\UseRawInputEncoding 

\documentclass[preprintnumbers, showpacs, floatfix,onecolumn,preprintnumbers, letterpaper, superscriptaddress,nofootinbib]{revtex4}
\usepackage{amsfonts}
\usepackage{amsmath}
\usepackage{amssymb,epsf}
\usepackage{latexsym}
\usepackage{graphicx,epsfig}
\usepackage{amssymb}
\usepackage{float}
\usepackage{subfigure}
\usepackage{epstopdf}
\usepackage[colorlinks=true,citecolor=blue,linkcolor=blue,urlcolor=black]{hyperref}
\usepackage{dcolumn}
\usepackage{psfrag}
\usepackage{wrapfig}
\usepackage{makeidx}                                          
\usepackage{epsf}

\begin{document}

\title{Non-minimal coupling Gauss-Bonnet holographic
superconductors}
\author{Mahya Mohammadi}
\email{mahya.mohammadi@hafez.shirazu.ac.ir}
 \affiliation{Department of Physics,
College of Sciences, Shiraz University, Shiraz 71454, Iran}
\affiliation{Biruni Observatory, College of Sciences, Shiraz
University, Shiraz 71454, Iran}

\author{Ahmad Sheykhi}
\email{asheykhi@shirazu.ac.ir}

 \affiliation{Department of Physics,
College of Sciences, Shiraz University, Shiraz 71454, Iran}
\affiliation{Biruni Observatory, College of Sciences, Shiraz
University, Shiraz 71454, Iran}

\begin{abstract}
By employing the gauge/gravity duality, we disclose the effects of
the non-minimal coupling between the scalar and Maxwell fields on
the holographic superconductors in the probe limit. As the
background spacetime, we consider a five dimensional Gauss-Bonnet
(GB) black hole with a flat horizon. We find out that the critical
temperature decreases for larger values of GB coupling constant,
$\alpha$, or smaller values of non-minimal coupling constant,
$\lambda$, which means that the condensation is harder to form. In
addition, we study the electrical conductivity in the holographic
setup. We show that at low frequency regime ($\omega\rightarrow
0$), Kramers-Kronig relation connects both parts of conductivity
to each other, while at high frequency
($\omega\rightarrow\infty$), two parts go up with constant slope.
Furthermore, we observe that the gap frequency shifts to larger
values for stronger $\alpha$, and becomes flat by increasing
$\lambda$.
\end{abstract}
\maketitle
\section{Introduction}
Condensed matter systems provide an area to test the concepts of
high energy theory experimentally. Within this field, there are
many strongly coupled systems, which are of significant
technological interest but challenging to observe \cite{1}. In
addition, there is no analytic way to compute the transport
coefficients, for such condensed matter or fluid systems directly.
One of transport coefficients of interest is the superconductivity
coefficient \cite{1}. Superconductivity is a phase of matter that
have no electrical resistance at low temperatures and since its
discovery, many researches have been done to understand its
physics \cite{2}. The BCS theory proposed by Bardeen, Cooper and
Schrieffer is almost an acceptable microscopic theory which
addresses the superconductivity as a microscopic effect originates
from condensation of Cooper pairs into a boson-like state through
the exchange of phonons \cite{3}. However, decoupling of Cooper
pairs at high temperatures is one of defects of this theory. The
Anti-de Sitter/Conformal Field Theory duality  known as AdS/CFT
correspondence may consider as an effective approach to describe
strong coupling systems through corresponding the strong coupling
conformal field theory living on the boundary in $d$-dimensions to
a weak coupling gravity in $(d+1)$-dimensional spacetime in the
bulk \cite{4}. One of the fascinating consequence of the
holographic approach is that instead of enumerating Boltzmann
partition sums for the thermal problem, we can solve a simple
black hole problem in the bulk. Within this approach, standard
thermodynamical quantities such as the free energy, thermal phase
diagrams and the entropy may be calculated \cite{1}. In addition,
the frequency-dependent conductivity, which is one of the
important observable characterizing the properties of condensed
matter systems, can be obtained in this approach \cite{1}. A very
instructive example of a quantum phase transition within
gauge/gravity duality is obtained by using a magnetic field as the
control parameter \cite{5,6,7,8,9,10}.

The holographic duality provides description of thermal properties
of matter, Fermi liquids, superfluids and superconductors. An
important characteristic of holographic models at finite charge
density is that there is a critical temperature below which a new
ground state with lower free energy forms. This new ground state
corresponds to a condensate \cite{9}. To introduce the temperature
in the holographic quantum field theory and to introduce a phase
transition, we need to find a black hole that has hair only at low
temperatures, and it has no hair at high temperatures \cite{11}.
On the other hand, Landau-Fermi liquid theory is a well-understood
approach for describing fermions in weakly coupled systems. These
systems posses a Fermi surface which contains essential
information about the physical properties of strongly coupled
systems. AdS/CFT duality provides a way to calculate spectral
functions and identifying Fermi surfaces for strongly coupled
systems \cite{12}. With time, lots of works have been done in the
context of holographic condensed matter. For instance in 2008,
Hartnoll and et al. presented the idea of holographic
superconductors\cite{14}. Many properties of high and
unconventional condensed matter systems such as free energy,
entropy, phase transition, electrical, thermal and
thermoelectrical conductivities and etc are obtained by applying
gauge/gravity duality \cite{13,15,16,18}. In spite of intense
researches in this subject (se e.g.
\cite{19,20,21,22,23,24,25,26,35,36,37,38,39,40,43,44,45,46,47,
49,50,51,52,53,54,55,56,57,58,59,60,61,62,63,64,65,66,66aa,66bb,66cc}),
there are still many open problems in strongly coupled systems
that each of them can help us to understand the physics of
strongly coupled condensed matter systems. Studying the effects of
non-minimal coupling on holographic condensed matter systems by
considering different forms of coupling is one of this problems.
Nowadays, scalarized black holes solutions have been attracted
much attention (see e.g. \cite{66a,66c,66d,66e}). Since the
non-minimal couplings can provide an effective mass for the scalar
field and lead to the spontaneous scalarization that can be
interpreted as the holography phase transition. In \cite{66f,me},
Einstein-Maxwell-scalar (EMS) model with a non minimal coupling
between the scalar and Maxwell fields were studied.

In this work, we study the effects of non-minimal coupling between
scalar and Maxwell fields in GB holographic superconductors. By
analyzing equations of motion numerically, we shall investigate
the influence of the GB coupling $\alpha$ and the non-minimal
coefficient $\lambda$ on the critical temperature $T_c$ and phase
transition of the system. In addition, by applying a suitable
electromagnetic perturbation on the black hole background, we
explore the behavior of real and imaginary parts of conductivity
for different values of $\alpha$ and $\lambda$.

This paper is organized as follows. In section \ref{sec2}, we
study holographic setup when scalar and Maxwell fields couple to
each other non-minimally in the background of GB gravity. In
section \ref{sec3}, we analyze the electrical conductivity in
holographic context. Finally, in section \ref{sec4}, we summarize
our results.
\section{The Holographic Model}\label{sec2}
In order to study the condensation of the scalar field $\psi$ with
mass $m$ and charge $q$ in the background of the AdS black holes
with a non-minimal coupling between the Maxwell and scalar fields
in the presence of GB gravity, we consider the action as
\cite{66f,66g,me}
\begin{eqnarray}\label{eqact}
&&S =\int d^{5}x\sqrt{-g} \left[\mathcal{L}_{G}+\mathcal{L}_{m}\right], \notag \\
&& \mathcal{L}_{G}= R-2 \Lambda+\frac{\alpha}{2}\left[R^{2}-4 R^{\mu\nu}
R_{\mu \nu}+R^{\mu \nu \rho \sigma} R_{\mu \nu \rho \sigma}\right] , \notag \\
&&\mathcal{L}_{m}=
-\frac{h(\psi)}{4} F_{\mu\nu}F^{\mu\nu}- \vert\nabla\psi-i q A \psi \vert^{2}
-m^{2} \vert \psi \vert^{2} ,\label{actgauss}
\end{eqnarray}
where $g$, $R$, $R_{\mu \nu}$ and $R_{\mu \nu \rho \sigma}$ are
metric determinant, Ricci scalar, Ricci tensor and Riemann
curvature tensor, respectively. $\Lambda$ represents the negative
cosmological constant and equals to $-6/l^{2}$ in five dimension
with $l$ as the radius of the AdS spacetime \cite{mahyalast}.
Here, we choose $l=1$. $\alpha$ is the GB parameter and in
$\alpha\rightarrow0$ limit, $ \mathcal{L}_{G}$ in Eq.
(\ref{eqact}) turns to Einstein case. In the above action,
$h(\psi)=1+\lambda \psi^2$ represents the non-minimal coupling
function with $\lambda\geq0$. We choose this form of $h(\psi)$ because it gives us more information about wide range of non-minimal coupling constant $\lambda$ and temperature \cite{me}.  By considering $A_{\mu}$ as the
vector potential, the strength of the Maxwell field is defined by
$F_{\mu\nu}=\nabla_{\mu}A_{\nu}-\nabla_{\nu}A_{\mu}$.

As the background spacetime, we consider a five dimensional GB
black hole which its geometry is given by the line elements
\cite{66g}
\begin{eqnarray} \label{metric2}
&&{ds}^{2}=-f(r){dt}^{2}+\frac{{dr}^{2}}{f(r)}+r^{2}({dx}^{2}+{dy}^{2}+{dz}^{2})%
,\\
&&f(r)=\frac{r^2}{2 \alpha } \left[1-\sqrt{1-4 \alpha  \left(1-\frac{1}{r^{4}}\right)}\right],\label{eqfgauss} %
\end{eqnarray}%
where the function $f(r)$ has the asymptotic behavior
$(r\rightarrow \infty)$ as
\begin{equation}
f(r)=\frac{ r^2}{2 \alpha }\left[1-\sqrt{1-4 \alpha }\right].
\end{equation}
We can present the effective radius $L_{\text{eff}}$ for the AdS
spacetime as \cite{23}
\begin{equation}
L_{\text{eff}}^2=\frac{2 \alpha }{1-\sqrt{1-4 \alpha }}.
\end{equation}
Varying action (\ref{eqact}) with respect to the scalar field
$\psi$ and the gauge field $A_{\mu}$, by choosing $\psi=\psi(r)$
and $A_{\mu}dx^{\mu}=\phi(r)dt$, yields to equations of motion
\begin{equation}\label{eqpsi}
\psi ''(r)+\left[\frac{f'(r)}{f(r)}+\frac{3}{r}\right]\psi '(r)+
\left[-\frac{m^2}{f(r)}+\frac{q^2 \phi (r)^2}{f(r)^2}+\frac{\lambda \phi '(r)^2}{2 f(r)}\right]\psi (r)=0,
\end{equation}
\begin{equation}\label{eqphi}
\phi ''(r)+ \left[\frac{2 \lambda  \psi (r) \psi '(r)}{(1+\lambda \psi (r)^2)}+\frac{3}{r}\right]\phi
'(r)-\frac{2 q^2 \psi (r)^2 }{f(r) \left(1+\lambda \psi (r)^2\right)}\phi
(r)=0.
\end{equation}
In $\lambda=0$, $\alpha=0$ and $\alpha=\lambda=0$ limits, we obtain the same equations as \cite{66g,me,2a}.

\begin{figure*}[t]
\centering
\subfigure[~$\alpha=-0.08$]{\includegraphics[width=0.4\textwidth]{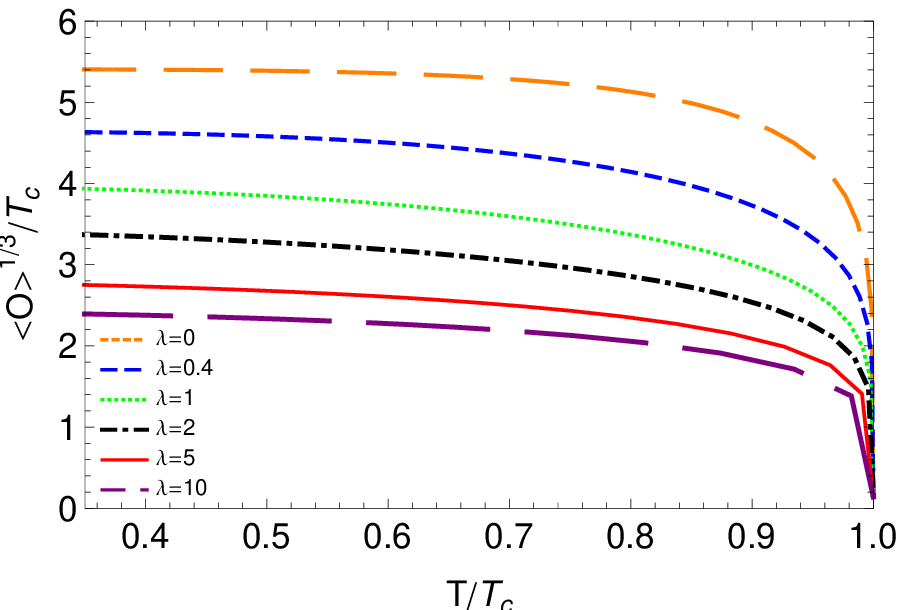}} \qquad %
\subfigure[~$\alpha=0.08$]{\includegraphics[width=0.4\textwidth]{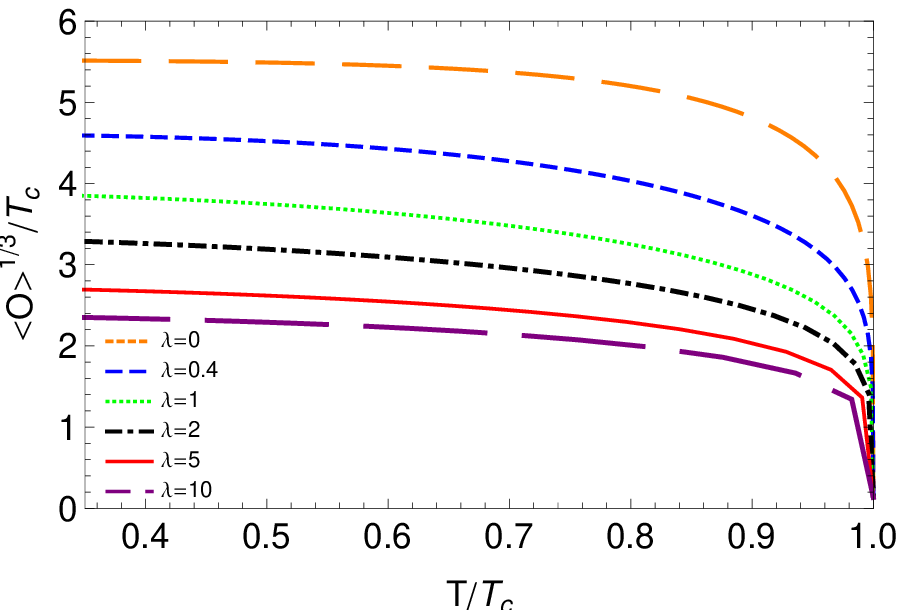}} \qquad %
\subfigure[~$\lambda=0$]{\includegraphics[width=0.4\textwidth]{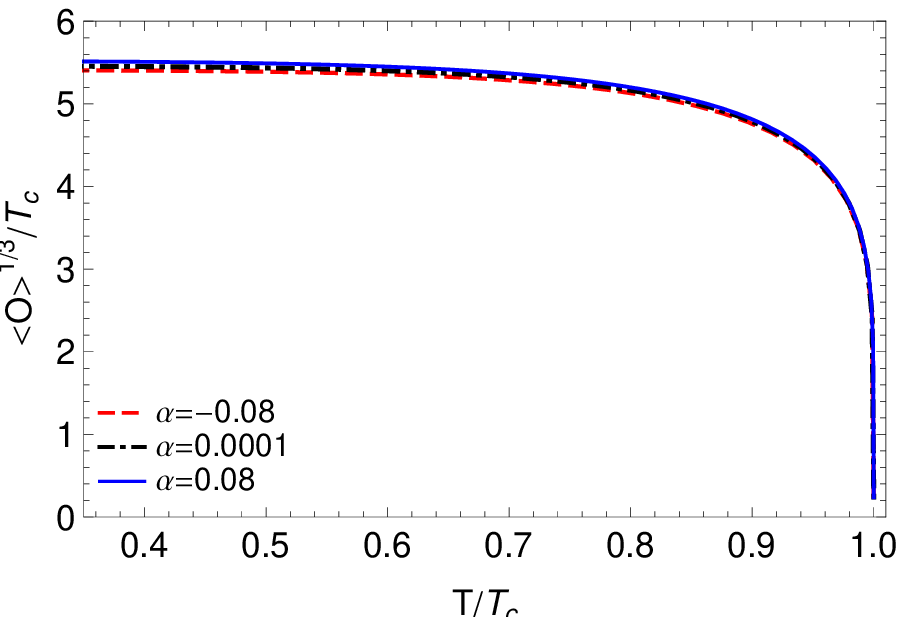}} \qquad %
\subfigure[~$\lambda=10$]{\includegraphics[width=0.4\textwidth]{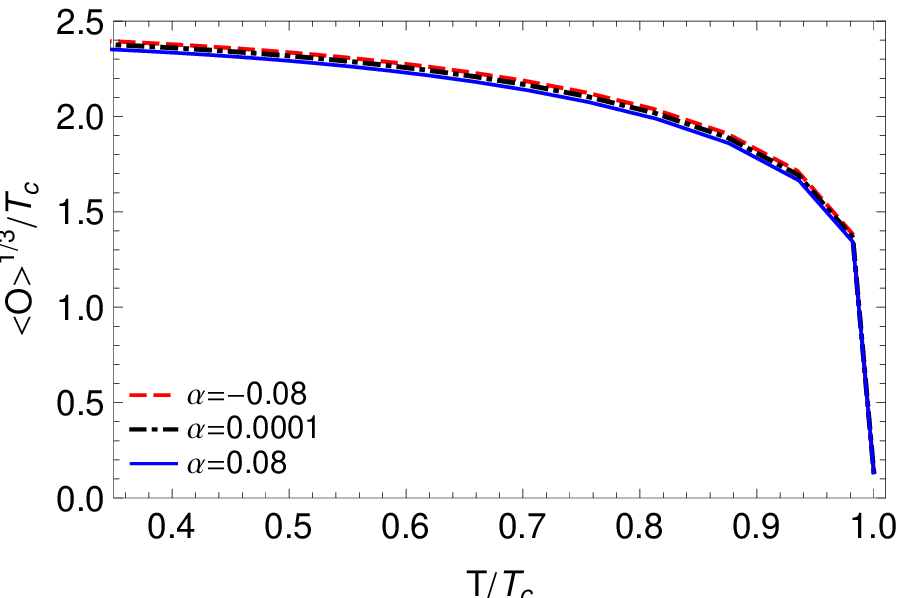}} \qquad %
\caption{The behavior of the condensation as a function of
temperature.}\label{fig1}
\end{figure*}

If we consider $\mu$ and $\rho$ as chemical potential and charge
density, the above equations have the asymptotic
($r\rightarrow\infty$) solutions as
\begin{equation}\label{eqasym}
\psi(r)=\frac{\psi_{+}}{r^{\Delta _+}}+\frac{\psi _{-}}{r^{\Delta _-}}, \  \  \ \phi(r) =\mu -\frac{\rho }{r^{2}},
\end{equation}
\begin{equation}\label{eqasym3}
\Delta _\pm=\left[2\pm\sqrt{4+ m^2 L_{\text{eff}}^{2}}\right],
\end{equation}

where the Breitenlohner-Freedman (BF) bound is given by
\begin{equation}
\overline{m}^{2}\geqslant -4, \  \   \    \ \overline{m}^{2}=m^{2} L_{\text{eff}}^{2}.
\end{equation}
In this work, we take $m^2=-3/L_{\text{eff}}^{2}$. Since we are looking for spontaneous
symmetry breaking, we set $\psi _{-}=0$ because it plays the role
of source. $\psi _{+}$ is known as expectation value of the order
parameter $\langle\mathcal{O_{+}}\rangle$. Without loss of
generality, we set $q=1$. Shooting method is a practical approach
to solve equations of motion numerically. With the help of this
method, we find the relation between critical temperature $T_{c}$
and $\rho^{1/3}$ for different values of $\lambda$ and $\alpha$.
The results are listed in table I and show that increasing the
effect of GB parameter $\alpha$ makes conductor/superconductor
phase transition more difficult. While larger values of $\lambda$
lead to higher critical temperature which is so interesting
because of the importance of high temperature superconductors. So,
this approach can provide a good theoretical way to explore the
trend of these matters at higher temperatures. In addition, the
behavior of condensation as a function of temperature for various
effects of $\lambda$ and $\alpha$ are shown in figure \ref{fig1}.
Larger values of condensation for stronger effect of $\alpha$ and
weaker effect of $\lambda$ are pointed the difficulty of phase
transition. In five dimension, Hawking temperature $T$ equals to
$r_{+}/\pi$ that $r_{+}$ defines the horizon location
\cite{mahyalast}. In this work, for simplicity we fix $r_{+}=1$.
\begin{table*}[t]
\label{tab5}
\begin{center}
\begin{tabular}{c|c|c|c|}
\cline{2-4}
& $\alpha=-0.08$ &$\alpha=0.0001$ & $\alpha=0.08$  \\
\hline
\multicolumn{1}{|c|}{$\lambda=0$} & $0.2022 $ $\rho^{1/3}$& $ 0.1980$ $\rho^{1/3}$ & $ 0.1928 $ $\rho^{1/3}$  \\
\hline
\multicolumn{1}{|c|}{$\lambda=0.4$} & $0.2228 $ $\rho^{1/3}$ & $ 0.2199$ $\rho^{1/3}$ & $ 0.2166$ $\rho^{1/3}$  \\
\hline
\multicolumn{1}{|c|}{$\lambda=0.8$} & $0.2399 $ $\rho^{1/3}$ & $ 0.2375$ $\rho^{1/3}$ & $ 0.2347$ $\rho^{1/3}$\\
\hline
\multicolumn{1}{|c|}{$\lambda=1$} & $0.2468 $ $\rho^{1/3}$ & $ 0.2445$ $\rho^{1/3}$ & $0.2419 $ $\rho^{1/3}$\\
\hline
\multicolumn{1}{|c|}{$\lambda=2$} & $0.2722 $ $\rho^{1/3}$ & $ 0.2700$ $\rho^{1/3}$ & $0.2675 $ $\rho^{1/3}$\\
\hline
\multicolumn{1}{|c|}{$\lambda=5$} & $0.3138 $ $\rho^{1/3}$ & $0.3115 $ $\rho^{1/3}$ & $0.3090 $ $\rho^{1/3}$\\
\hline
\multicolumn{1}{|c|}{$\lambda=10$} & $0.3510 $ $\rho^{1/3}$ & $0.3485 $ $\rho^{1/3}$ & $0.3458 $ $\rho^{1/3}$\\
\hline
\end{tabular}%
\caption{Numerical results of critical temperature for different
values of GB parameters $\alpha$ and non-minimal coupling term
$\lambda$.}
\end{center}
\end{table*}
\section{Conductivity}\label{sec3}
\begin{figure*}[t]
\centering
\subfigure[~$\alpha=-0.08$, $\lambda=0$]{\includegraphics[width=0.4\textwidth]{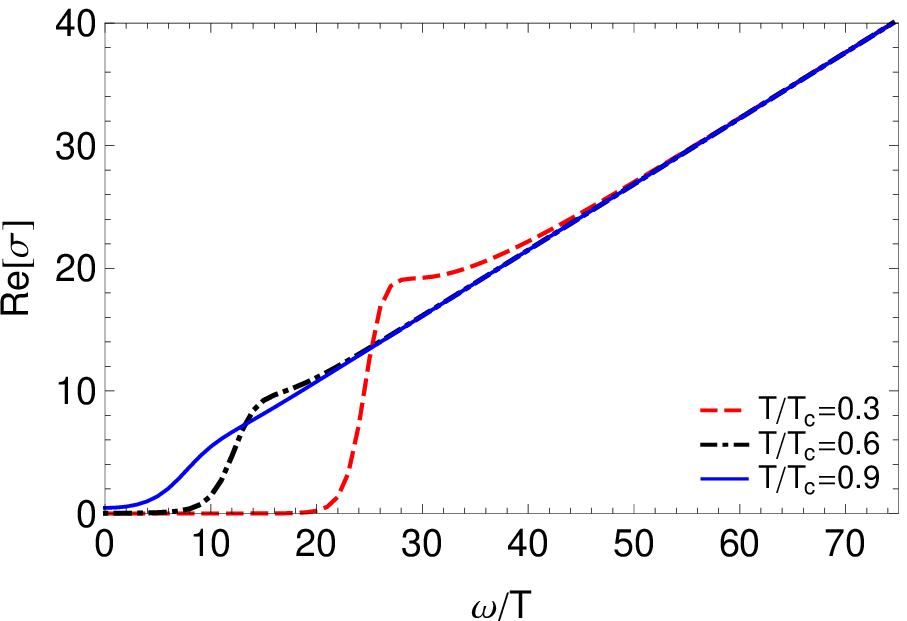}} \qquad %
\subfigure[~$\alpha=-0.08$, $\lambda=10$]{\includegraphics[width=0.4\textwidth]{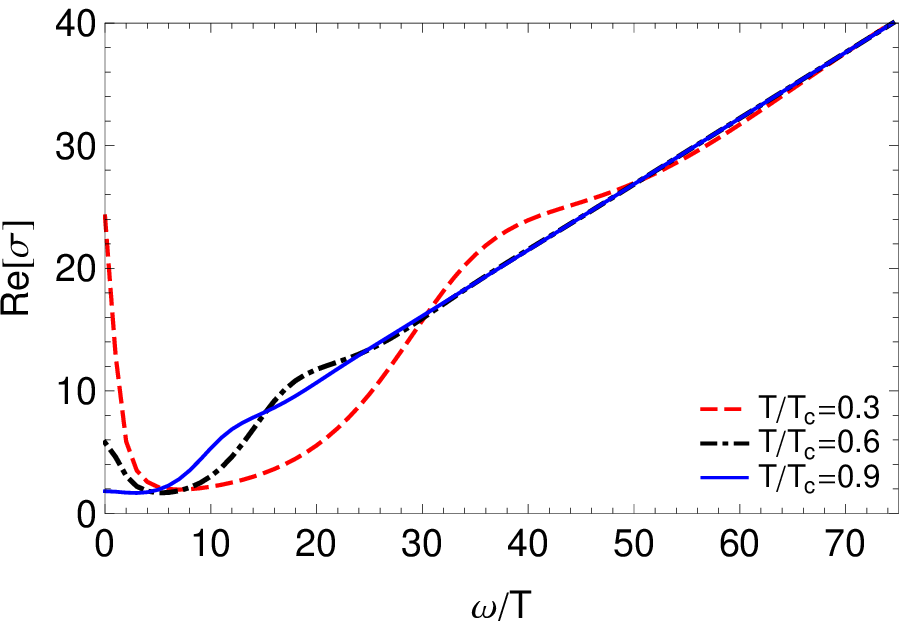}} \qquad %
\subfigure[~$\alpha=0.08$, $\lambda=0$]{\includegraphics[width=0.4\textwidth]{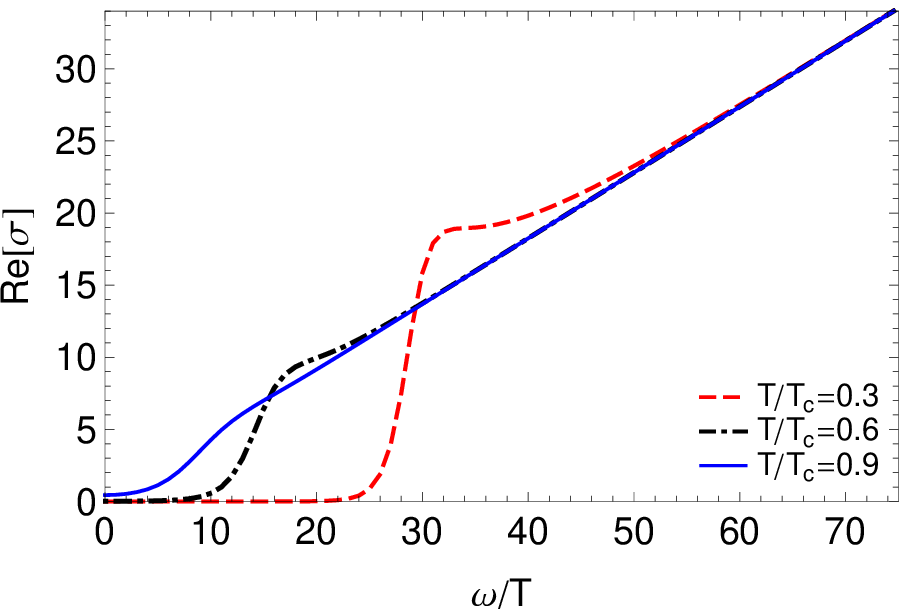}} \qquad %
\subfigure[~$\alpha=0.08$, $\lambda=10$]{\includegraphics[width=0.4\textwidth]{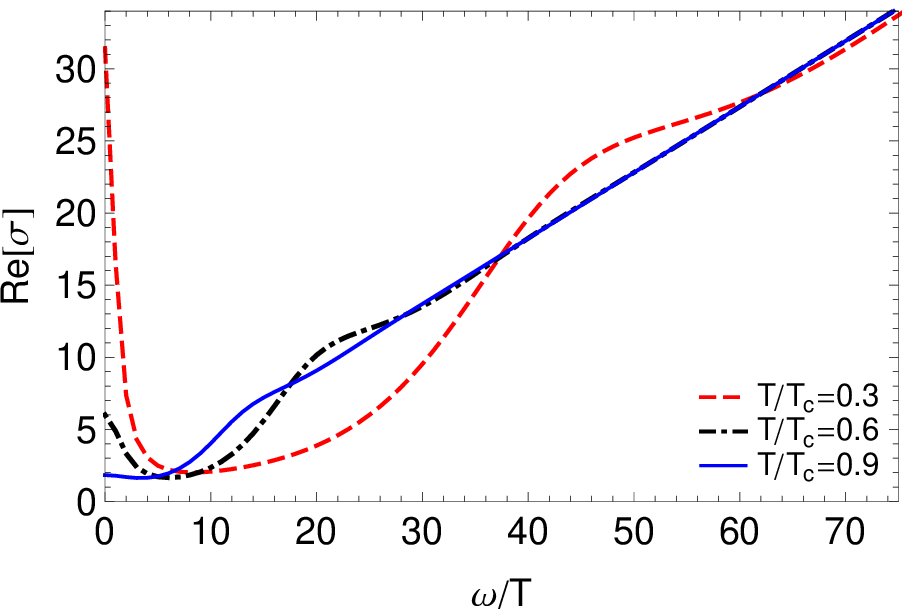}} \qquad %

\caption{The behavior of real parts of conductivity}\label{fig2}
\end{figure*}

\begin{figure*}[t]\label{fig3}
\centering
\subfigure[~$\alpha=-0.08$, $\lambda=0$]{\includegraphics[width=0.4\textwidth]{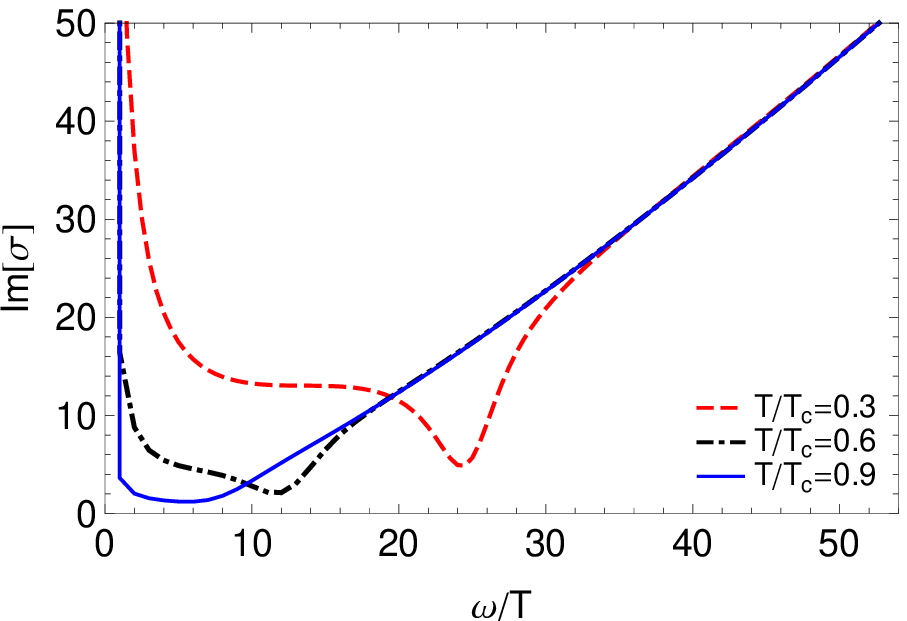}} \qquad %
\subfigure[~$\alpha=-0.08$, $\lambda=10$]{\includegraphics[width=0.4\textwidth]{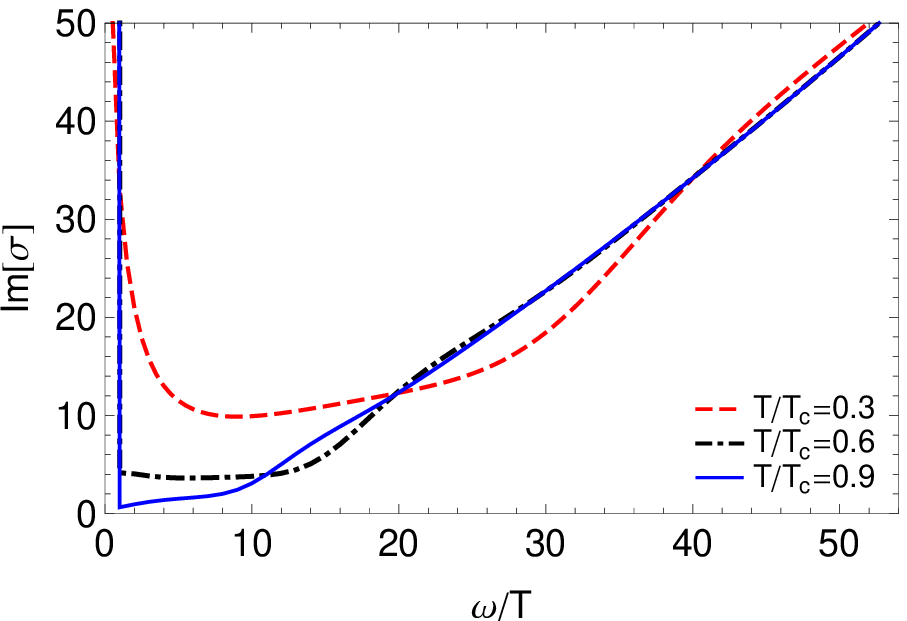}} \qquad %
\subfigure[~$\alpha=0.08$, $\lambda=0$]{\includegraphics[width=0.4\textwidth]{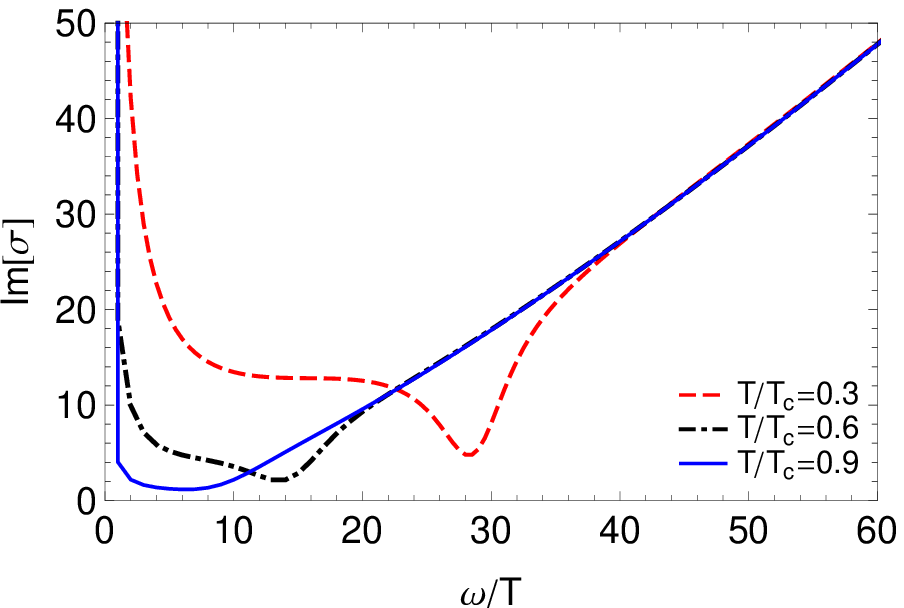}} \qquad %
\subfigure[~$\alpha=0.08$, $\lambda=10$]{\includegraphics[width=0.4\textwidth]{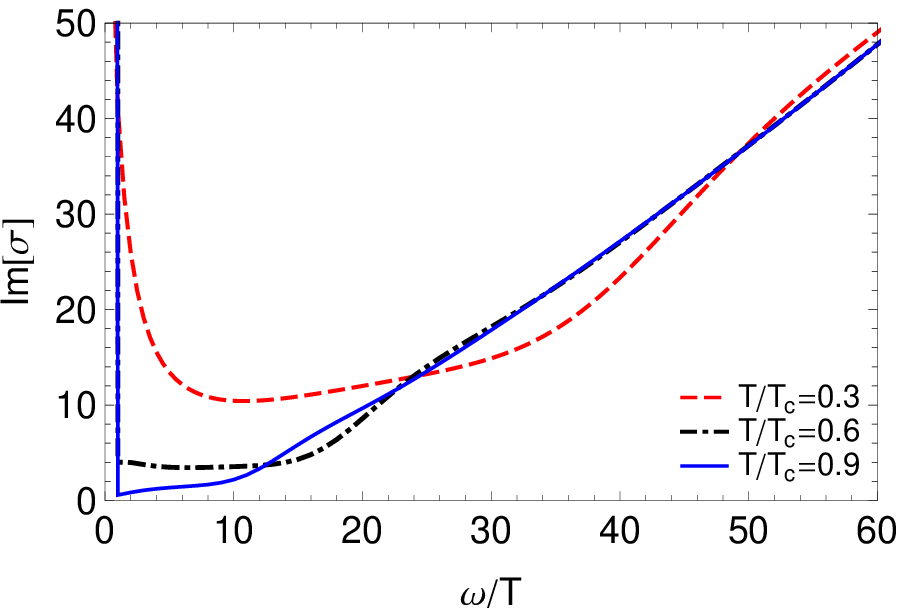}} \qquad %

\caption{The behavior of imaginary parts of conductivity}
\end{figure*}

\begin{figure*}[t]
\centering
{\includegraphics[width=0.4\textwidth]{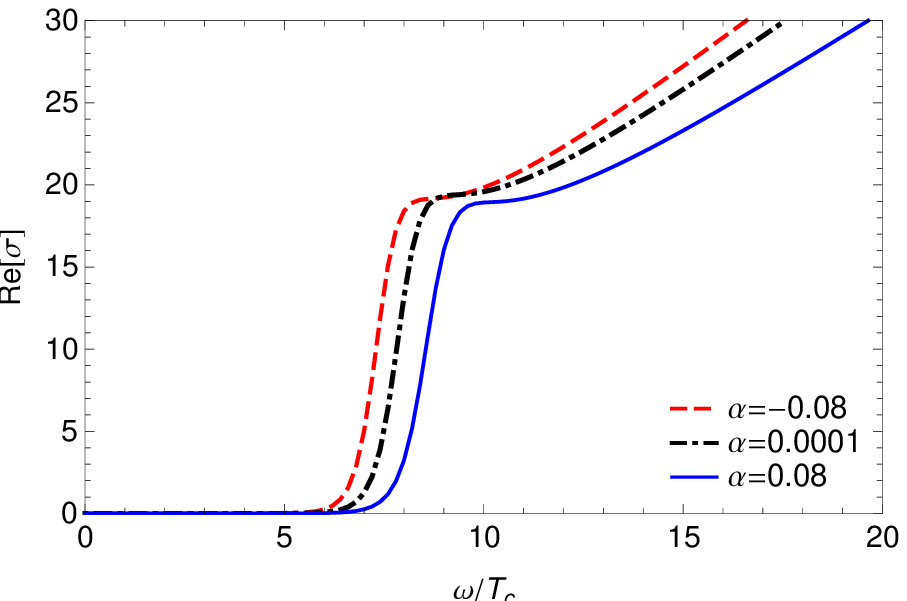}} \qquad %
{\includegraphics[width=0.4\textwidth]{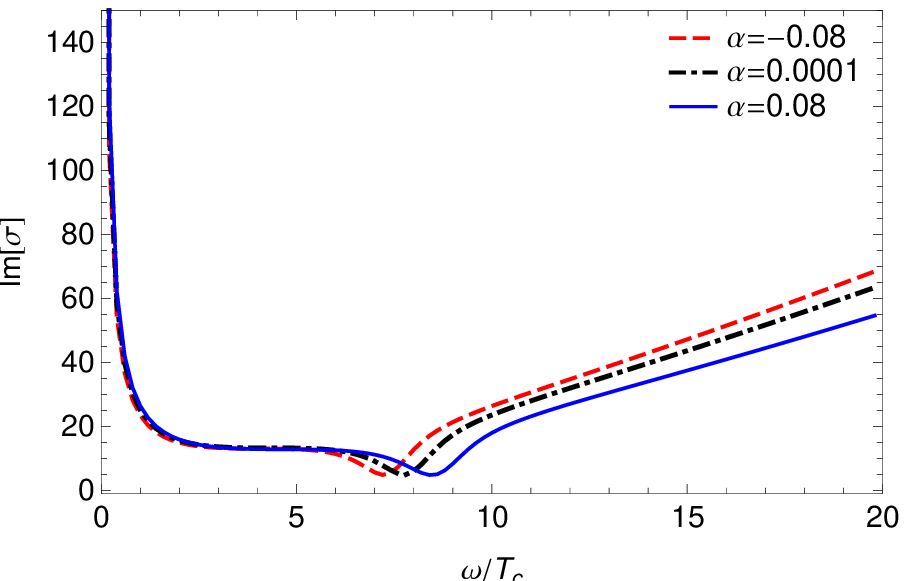}} \qquad %
\caption{The behavior of real and imaginary parts of conductivity
for $\lambda=0$ in $T/T_{c}=0.3$ }\label{fig4}
\end{figure*}

\begin{figure*}[t]\label{fig5}
\centering
{\includegraphics[width=0.4\textwidth]{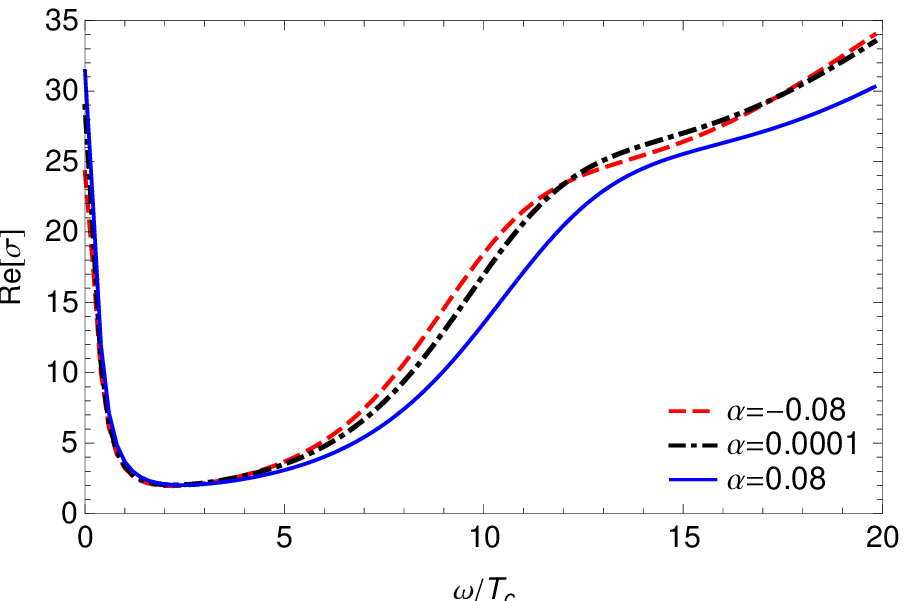}} \qquad %
{\includegraphics[width=0.4\textwidth]{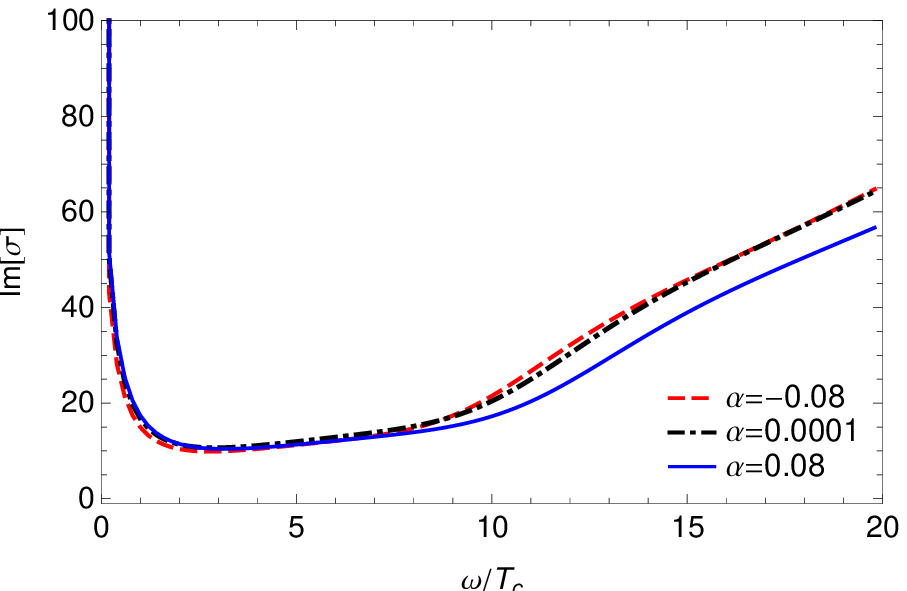}} \qquad %
\caption{The behavior of real and imaginary parts of conductivity
for $\lambda=10$ in $T/T_{c}=0.3$ }
\end{figure*}

\begin{figure*}[t]\label{fig6}
\centering
{\includegraphics[width=0.4\textwidth]{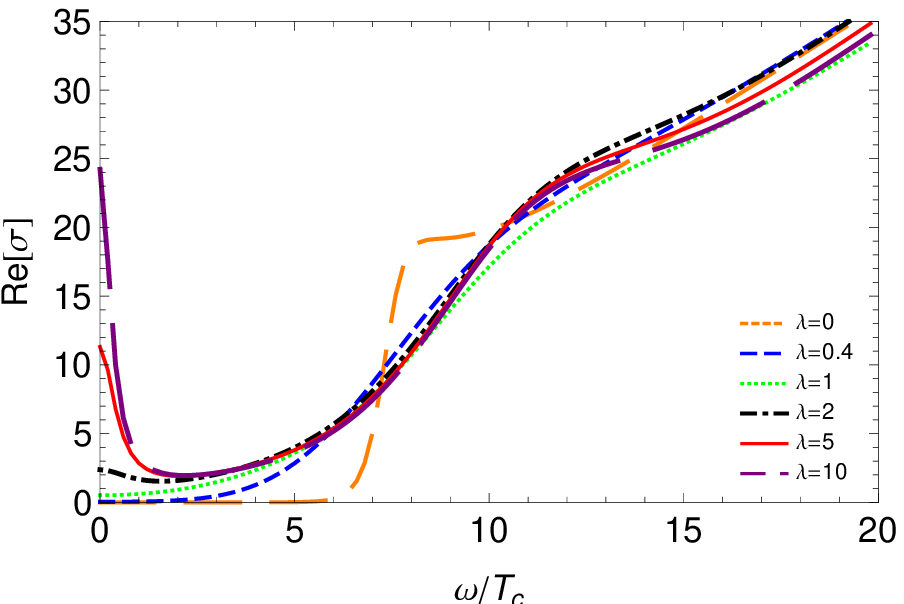}} \qquad %
{\includegraphics[width=0.4\textwidth]{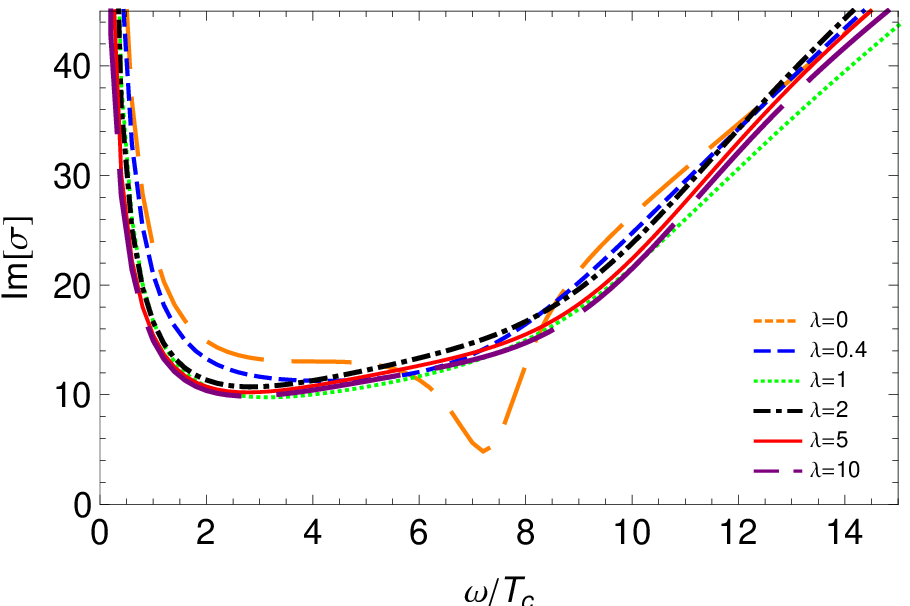}} \qquad %
\caption{The behavior of real and imaginary parts of conductivity
for $\alpha=-0.08$ in $T/T_{c}=0.3$ }
\end{figure*}

\begin{figure*}[t]
\centering
{\includegraphics[width=0.4\textwidth]{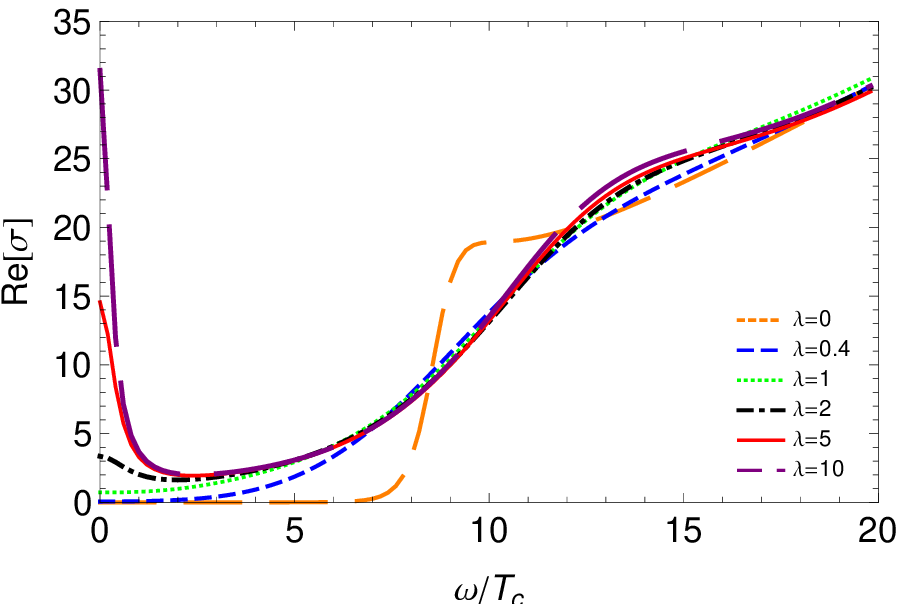}} \qquad %
{\includegraphics[width=0.4\textwidth]{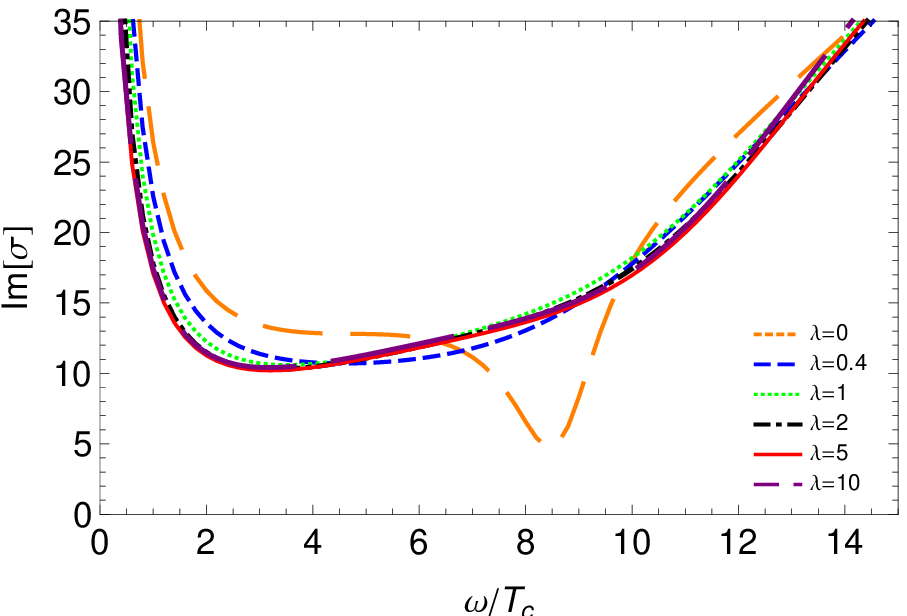}} \qquad %
\caption{The behavior of real and imaginary parts of conductivity
for $\alpha=0.08$ in $T/T_{c}=0.3$ }\label{fig7}
\end{figure*}

In this section, we are going to calculate electrical conductivity
in holographic setup through turning on an appropriate
electromagnetic perturbation as $\delta A_x = A_x e^{-i \omega t}$
on the black hole background which is dual to the boundary
electrical current. Turning on this component yields to
\begin{equation}
A_{x}''(r)+ \left[\frac{1}{r}+\frac{f'(r)}{f(r)}+\frac{2 \lambda \psi (r) \psi '(r)}{(1+\lambda \psi (r)^2)}\right]A_{x}'(r)+
\left[\frac{\omega ^2}{f(r)^2}-\frac{2 q^2 \psi (r)^2}{f(r) (1+\lambda \psi (r)^2)}\right]A_{x}(r)=0,
\end{equation}
which behaves asymptotically as
\begin{equation}
A_{x}''(r)+\frac{3}{r}A_{x}'(r)+\frac{\omega ^2  L_{\text{eff}}^4}{r^4}A_{x}(r)=0,
\end{equation}
by considering $A^{(0)}$ and $A^{(1)}$ as constant parameters, the
asymptotic solution of $A_x$ is
\begin{equation} \label{aysolgauss}
A_{x} =A^{(0)}+\frac{A^{(1)}}{r^2}+\frac{ \omega ^2
L_{\text{eff}}^4 \log (\Lambda  r)}{2 r^2}A^{(0)}+\cdots.
\end{equation}%
Thus, the electrical conductivity based on holographic approach
has the following form
\begin{equation} \label{conductivitygauss}
\sigma_{xx} =\frac{2 A^{(1)}}{i \omega L_{\text{eff}}^2 A^{(0)}}+\frac{i \omega L_{\text{eff}}^2}{2},
\end{equation}
By applying the ingoing wave boundary condition, the behavior of
conductivity as a function of frequency are shown in figures
\ref{fig2}-\ref{fig7}. Based on these figures, at low frequency
regime $\omega\rightarrow 0$, the real and imaginary parts of
conductivity are connected to each other with Kramers-Kronig
relation by showing a delta behaviour and pole respectively. on
the other hand, when $\omega\rightarrow\infty$ both parts increase
with constant slope which is proportional to $\omega$.
Furthermore, superconducting gap is occurred at temperatures below
the critical value and becomes sharper at lower temperatures which
shows difficulty of conductor/superconductor phase transition. To
illustrate the effects of $\alpha$ and $\lambda$ on the gap
frequency, we plot the trend of real and imaginary parts at fixed
temperature $T=0.3 T_c$ in figures \ref{fig4}-\ref{fig7}. It is
seen that $\omega_g$ shifts to larger values for stronger effect
of GB parameter which proves the results of previous section.
In addition, we face with flatter gap for larger $\lambda$ values.
Based on the BCS theory $\omega_g \simeq 3.5 T_c$, but in
holographic approach due to strong coupling between the pairs,
$\omega_g \simeq 8 T_c$.

\section{Summary and discussion}\label{sec4}
By applying AdS/CFT duality, we explored the holographic
superconductors in the context of GB gravity when the Maxwell and
scalar fields couple to each other non-minimally. First, we
investigated the effects of different values of $\alpha$ and
$\lambda$ on the critical temperature and condensation by solving
the equations of motion numerically in the probe limit. We face
with lower values of critical temperature $T_c$ and higher values of condensation
by increasing $\alpha$ or decreasing $\lambda$ which indicates the
difficulty of the conductor/superconductor phase transitions.
Next, we analyzed the behavior of real and imaginary parts of
conductivity through turning on an appropriate electromagnetic
perturbation on black hole back ground. Based on our results,
Kramers-Kronig relation connects these two parts at low frequency
regime while in $\omega\rightarrow\infty$ limit both parts of
conductivity go up with a constant slope proportional to $\omega$.
By decreasing temperature below the critical value, the gap
frequency occurs at about $\omega_g\simeq 8 T_c$ which shifts to
higher frequencies by decreasing temperature or increasing the GB
coupling constant $\alpha$. Stronger effects of non-minimal
coupling term $\lambda$ make this gap flatter.


\end{document}